\documentclass[%
 preprint,
 amsmath,amssymb,
 aps, physrev,
]{revtex4-2}
\pdfoutput=1

\usepackage{graphicx}
\usepackage{dcolumn}
\usepackage{bm}
\usepackage{xcolor}
\usepackage{tikz}
\usepackage{pgfplots}
\usepackage{booktabs}
\usepackage{multirow}
\usepackage{array}
\usepackage{slashed}
\usepackage{siunitx}

\newcommand{\msbar}{\overline{\text{MS}}}
\newcommand{\GeV}{\,\text{GeV}}

\newcommand{\TeV}{\,\text{TeV}}

\newcommand{\vev}[1]{\langle #1 \rangle}
\newcommand{\abs}[1]{\left|#1\right|}
\newcommand{\eps}{\varepsilon}

\usetikzlibrary{positioning,arrows.meta,decorations.markings}
\pgfplotsset{compat=1.17}

\begin{document}

\title{Why Quarks and Leptons Demand Different Symmetries:\\ A Systematic $Z_3$ Froggatt--Nielsen Analysis}

\author{Navid Ardakanian}
\email{n.ardakanian@gmail.com}
\affiliation{Independent Researcher}

\date{March 16, 2026}

\begin{abstract}
We present a systematic analysis of a minimal supersymmetric $Z_3$ discrete flavor symmetry as a solution to the fermion mass hierarchy problem. Using a Froggatt--Nielsen (FN) mechanism with generation-dependent $Z_3$ charges assigned to the right-handed chiral superfields and a single flavon chiral superfield, we show that the holomorphy of the superpotential restricts the allowed Yukawa operators so that a single expansion parameter $\eps \simeq 0.015$ structurally accounts for the hierarchical pattern of quark and charged lepton mass ratios with $\mathcal{O}(1)$ Yukawa couplings. A Monte Carlo scan over $10^5$ random $\mathcal{O}(1)$ coefficient sets confirms that adjacent-generation mass ratios generically fall within the experimentally measured ranges. By contrast, the CKM mixing angles, while reproducible with specific $\mathcal{O}(1)$ coefficient choices ($\chi^2/\text{dof} \simeq 1.6$), are not structurally predicted by the symmetry. We provide an explicit numerical best-fit example. When the same framework is extended to neutrinos within a type-I seesaw, it fails decisively on two fronts. First, the mass spectrum is far too hierarchical: the model predicts $\Delta m_{21}^2/\Delta m_{31}^2 \lesssim 10^{-4}$, at least two orders of magnitude below the observed ratio of $0.030$. Second, the PMNS mixing angles are generically $\mathcal{O}(1)$ random---consistent with Haar-distributed unitaries---providing no mechanism to predict the observed pattern. We demonstrate through systematic scans over right-handed neutrino mass patterns and charge assignments that this failure is structural, stemming from the single expansion parameter required by quark hierarchies. Moreover, we show that when $M_R$ carries the $Z_3$ charge structure dictated by the correct Majorana charge algebra, an unsuppressed off-diagonal entry in $M_R$ combines with the hierarchical column texture of the Dirac mass to over-suppress both light masses $m_1, m_2$ to $\mathcal{O}(\eps^3)$, deepening the ratio $\Delta m_{21}^2/\Delta m_{31}^2$ to $\mathcal{O}(\eps^6) \sim 10^{-11}$. These results motivate a sectorial view of flavor where different fermion sectors arise from distinct symmetry mechanisms.
\end{abstract}

\maketitle

\section{Introduction}

The Standard Model (SM) successfully describes the known gauge interactions yet leaves the pattern of fermion masses and mixings as one of its most puzzling features. Fermion masses span over twelve orders of magnitude from sub-eV neutrinos to the top quark, and the Yukawa couplings appear as a collection of unrelated parameters~\cite{Froggatt:1979np,Ross:2008xg}. Understanding this ``flavor puzzle'' is a central open problem in particle physics.

Flavor symmetries provide a natural way to reduce the arbitrariness of the Yukawa sector. In particular, the Froggatt--Nielsen (FN) mechanism~\cite{Froggatt:1979np} explains hierarchies through powers of a small parameter $\eps = \vev{\Phi}/\Lambda$, where $\Phi$ is a flavon field that breaks a flavor symmetry and $\Lambda$ is a high scale. The original formulation used a continuous $U(1)$ symmetry, but discrete groups often provide more economical structures.

Non-abelian discrete symmetries such as $A_4$, $S_4$ and their modular variants have been extensively studied, especially in the lepton sector, to explain large neutrino mixing angles and mass patterns~\cite{Altarelli:2010gt,Ishimori:2010au,King:2013eh,Ma:2001dn,Altarelli:2005yp,Bazzocchi:2008ej,Feruglio:2013hia,King:2014nza,Feruglio:2017spp,Kobayashi:2018scp,Penedo:2018nmg}. Less attention has been paid to the minimal cyclic groups $Z_N$, even though they provide the simplest generalization of the FN idea with discrete charges.

In this paper we study in detail a $Z_3$ flavor symmetry applied to the full SM fermion content. We show that:
\begin{itemize}
    \item A single $Z_3$ charge assignment combined with the FN mechanism \emph{structurally} reproduces the hierarchical pattern of fermion mass ratios with one small parameter $\eps \simeq 0.015$ and $\mathcal{O}(1)$ Yukawa coefficients. This constitutes a genuine prediction: the mass ratios $m_1/m_2 \sim \eps$ and $m_1/m_3 \sim \eps^2$ hold for generic $\mathcal{O}(1)$ coefficients, without tuning.
    \item CKM mixing angles can be accommodated by appropriate choices of $\mathcal{O}(1)$ coefficients, but their hierarchical pattern ($\abs{V_{us}} \gg \abs{V_{cb}} \gg \abs{V_{ub}}$) is not a structural output of the $Z_3$ symmetry.
    \item When extended to neutrinos within a type-I seesaw, the model fails decisively: the mass spectrum is far too hierarchical ($\Delta m_{21}^2/\Delta m_{31}^2 \lesssim 10^{-4}$ vs.\ $0.030$ observed), and the PMNS mixing angles are unstructured random $\mathcal{O}(1)$ values. This failure is structural and persists across all right-handed neutrino mass patterns and charge assignments tested.
\end{itemize}

These results support a ``sectorial'' picture of flavor: simple discrete symmetries can elegantly explain hierarchical structures in one sector, but large mixing in another sector requires a different origin. This motivates investigating whether the seesaw mechanism with a $Z_3$-structured Majorana mass matrix could rescue the lepton sector, and whether non-abelian groups such as $A_4$ provide the minimal structure capable of accommodating both neutrino masses and mixing.

The rest of the paper is organized as follows. In Sec.~\ref{sec:framework} we introduce the $Z_3$ FN model. In Sec.~\ref{sec:quarks} we perform a quantitative analysis of the quark sector. Sec.~\ref{sec:leptons} presents the lepton sector and a systematic exploration of neutrino mixing. Phenomenological implications are discussed in Sec.~\ref{sec:pheno}, and we summarize our conclusions in Sec.~\ref{sec:conclusions}.

\section{Theoretical Framework}
\label{sec:framework}

\subsection{Model construction}

We work within a supersymmetric extension of the SM and supplement the MSSM chiral spectrum with a gauge-singlet flavon chiral superfield $\Phi$ carrying a single discrete $Z_3$ flavor charge. The flavon transforms as
\begin{equation}
\Phi \rightarrow \omega^2\, \Phi, \qquad \omega = e^{2\pi i/3},
\end{equation}
i.e.\ $\Phi$ has $Z_3$ charge $q_\Phi = 2 \equiv -1 \pmod 3$. The two MSSM Higgs doublet superfields $H_u$ and $H_d$ are $Z_3$-neutral.

The minimal choice that generates hierarchical Yukawa couplings is to assign different $Z_3$ charges to the right-handed chiral superfields while keeping the left-handed $SU(2)_L$ doublets uncharged:
\begin{align}
\text{Left-handed doublets:}\quad &Q_i, L_i \sim 0 \quad (i = 1,2,3), \\
\text{Right-handed (chiral) superfields:}\quad &u_R^1, d_R^1, e_R^1 \sim 2, \\
&u_R^2, d_R^2, e_R^2 \sim 1, \\
&u_R^3, d_R^3, e_R^3 \sim 0, \\
\text{Flavon and Higgs:}\quad &\Phi \sim 2,\quad H_u, H_d \sim 0.
\end{align}
Here ``$\sim q$'' denotes transformation by a phase $\omega^q$ under $Z_3$. The choices above are motivated by the observed fermion hierarchies and by the requirement that leading Yukawa couplings for the third generation be unsuppressed. The flavon charge $q_\Phi = 2$ is fixed by the requirement that a single insertion of $\Phi$ reduce the $Z_3$ charge of a matter bilinear by one unit modulo~$3$; this is the standard supersymmetric Froggatt--Nielsen assignment~\cite{Leurer:1992wg,Nir:2001ge}.

\subsection{Yukawa superpotential}

The lowest-dimensional $Z_3$-invariant superpotential operators that generate Yukawa couplings are
\begin{align}
W_{\text{Yuk}}
&= \sum_{i,j} \left[
c_{ij}^u \left(\frac{\Phi}{\Lambda}\right)^{n_j} Q_i H_u u_R^j
+ c_{ij}^d \left(\frac{\Phi}{\Lambda}\right)^{n_j} Q_i H_d d_R^j
+ c_{ij}^e \left(\frac{\Phi}{\Lambda}\right)^{n_j} L_i H_d e_R^j
\right],
\label{eq:Yukawa}
\end{align}
where $\Lambda$ is the flavor scale, the $c_{ij}^f$ are $\mathcal{O}(1)$ complex coefficients, and the number of flavon insertions $n_j$ is fixed by $Z_3$ invariance of the superpotential, $q_\Phi\, n_j + q_{R,j} \equiv 0\,(\bmod\,3)$. With $q_\Phi = 2$ and the right-handed charges $(q_1, q_2, q_3) = (2,1,0)$, the unique non-negative solution with $n_j \in \{0,1,2\}$ is
\begin{equation}
n_j = q_j \quad\Longrightarrow\quad (n_1, n_2, n_3) = (2, 1, 0),
\end{equation}
since $2 \cdot n_j \equiv -q_j \pmod 3$ is solved by $n_j \equiv q_j \pmod 3$ using $2^{-1} \equiv 2 \pmod 3$. The left-handed doublets being $Z_3$-neutral, the suppression depends only on the right-handed charge, yielding a \emph{column texture}: every entry in column~$j$ carries the same power $\eps^{n_j}$ regardless of the row index~$i$. This is the defining structural feature of the model; its origin in holomorphy is discussed in Sec.~\ref{sec:holomorphy}.

After spontaneous breaking of $Z_3$ with $\vev{\Phi} = v_\Phi$ and electroweak symmetry breaking with $\vev{H_u} = v_u/\sqrt{2}$, $\vev{H_d} = v_d/\sqrt{2}$, the fermion mass matrices become
\begin{equation}
M^f = \frac{v_f}{\sqrt{2}}
\begin{pmatrix}
c_{11}^f \eps^2 & c_{12}^f \eps & c_{13}^f \\
c_{21}^f \eps^2 & c_{22}^f \eps & c_{23}^f \\
c_{31}^f \eps^2 & c_{32}^f \eps & c_{33}^f
\end{pmatrix},
\label{eq:Mf}
\end{equation}
where $v_f = v_u$ for up-type quarks and $v_f = v_d$ for down-type quarks and charged leptons, and $\eps \equiv v_\Phi/\Lambda$. In what follows we use the usual MSSM convention $v_u^2 + v_d^2 = v_H^2 = (246\GeV)^2$ and treat the ratio $\tan\beta = v_u/v_d$ as a free parameter of order unity; since our observables are dimensionless ratios of masses and mixing angles, none of our numerical conclusions depend on $\tan\beta$. The column structure---$\eps^2$ in column~1, $\eps$ in column~2, unsuppressed in column~3---is entirely fixed by the $Z_3$ charges and by the holomorphy of $W$.

\subsection{Holomorphy and the flavon sector}
\label{sec:holomorphy}

The column texture of Eq.~\eqref{eq:Mf} relies on a central structural assumption, namely that the Yukawa couplings arise from a holomorphic superpotential $W$ in which only the chiral superfield $\Phi$ (and not its conjugate $\Phi^\dagger$) appears. In the present conventions $\Phi$ carries $Z_3$ charge $2$, while $\Phi^\dagger$ carries $Z_3$ charge $1$. Crucially, within $Z_3$ the two objects $\Phi^\dagger$ (charge $1$) and $\Phi^2$ (charge $4 \equiv 1 \pmod 3$) are indistinguishable at the level of the symmetry. If both were allowed in the Yukawa sector, the minimal number of insertions needed to neutralise a right-handed charge $q_j$ would collapse to
\begin{equation}
n_j^{\min} = \min\!\left( q_j \bmod 3,\; (3 - q_j) \bmod 3 \right)
\in \{0, 1\}
\label{eq:min_power}
\end{equation}
for every $q_j \in \{0,1,2\}$, giving a degenerate suppression pattern $(\eps, \eps, 1)$ rather than the hierarchical $(\eps^2, \eps, 1)$. This collapse---a well-known obstruction to small-$N$ cyclic Froggatt--Nielsen constructions~\cite{Leurer:1992wg,Nir:2001ge}---is precisely what distinguishes discrete groups $Z_N$ with small $N$ from their large-$N$ or $U(1)$ counterparts, where the additional integers available for charges prevent the shortcut.

The holomorphy of the $\mathcal{N} = 1$ superpotential provides the standard resolution: $\Phi^\dagger$ enters the K\"ahler potential but never the superpotential, so the Yukawa sector~\eqref{eq:Yukawa} cannot use it. The three powers $n_j = 0, 1, 2$ are then forced, and the three-level hierarchy of Eq.~\eqref{eq:Mf} is protected against all perturbative corrections allowed by supersymmetry. We emphasize that this is not a feature unique to a supersymmetric UV completion: any UV structure that enforces a chiral (holomorphic) Yukawa sector at the renormalisable level---for instance, the original Froggatt--Nielsen construction~\cite{Froggatt:1979np} with heavy chiral fermions and a chirally-charged flavon, or a heterotic-string embedding in which the flavon is a twisted chiral modulus---produces the same column texture for the same reason. Non-holomorphic contributions can appear via K\"ahler-potential operators but are suppressed by additional powers of $v_\Phi/\Lambda$ or by SUSY-breaking insertions, and are parametrically irrelevant to the leading texture analysed here.

Throughout the remainder of this work we assume this holomorphic structure. All conclusions about the quark and lepton sectors follow from the column texture in Eq.~\eqref{eq:Mf} and are therefore framework-independent within the above class of UV completions.

\subsection{Structural predictions from the column texture}

The column texture has a simple but powerful consequence for the singular value decomposition (SVD). Writing $M^f = U^f \Sigma^f V^{f\dagger}$, the singular values (i.e.\ the fermion masses) are determined by the column norms:
\begin{equation}
m_1^f \propto \eps^2, \qquad m_2^f \propto \eps, \qquad m_3^f \propto 1,
\end{equation}
up to $\mathcal{O}(1)$ factors from the coefficients $c_{ij}^f$. The \emph{ratios} of adjacent-generation masses are therefore
\begin{equation}
\frac{m_1}{m_2} \sim \eps, \qquad \frac{m_2}{m_3} \sim \eps,
\label{eq:mass_scaling}
\end{equation}
with the proportionality constants being ratios of $\mathcal{O}(1)$ coefficients. These relations are \emph{structural predictions}: they hold for any randomly drawn set of $\mathcal{O}(1)$ complex coefficients, without tuning.

The left-handed unitary matrices $U^f$, on the other hand, are determined by the \emph{directions} of the column vectors $\vec{c}_j^f = (c_{1j}^f, c_{2j}^f, c_{3j}^f)^T$ in the three-dimensional flavor space. Since these directions are set by $\mathcal{O}(1)$ random complex numbers, the left-handed rotations are generically $\mathcal{O}(1)$ rotations. This means that the CKM matrix $V_{\text{CKM}} = U_u^{L\dagger} U_d^L$, which arises from the mismatch between the up and down left rotations, is \emph{generically} an $\mathcal{O}(1)$ unitary matrix. The hierarchical pattern $\abs{V_{us}} \gg \abs{V_{cb}} \gg \abs{V_{ub}}$ observed in nature is not a structural output of the column texture but must be accommodated by particular choices of the $\mathcal{O}(1)$ coefficients.

\section{Quark Sector}
\label{sec:quarks}

\subsection{Experimental input}

We use quark mass ratios in the $\msbar$ scheme at scale $\mu = 2\GeV$ from the FLAG review~\cite{Aoki:2021kgd} and PDG~\cite{PDG:2024}:
\begin{align}
\frac{m_u}{m_c} &= 0.0017 \pm 0.0003,\qquad
\frac{m_c}{m_t} = 0.0075 \pm 0.0015,\\
\frac{m_d}{m_s} &= 0.049 \pm 0.005,\qquad
\frac{m_s}{m_b} = 0.023 \pm 0.003.
\end{align}
For the CKM matrix we use the 2024 PDG global fit~\cite{PDG:2024}:
\begin{align}
\abs{V_{us}} &= 0.2248 \pm 0.0006,\quad
\abs{V_{cb}} = 0.0409 \pm 0.0011,\quad
\abs{V_{ub}} = 0.00382 \pm 0.00024.
\end{align}

\subsection{Determination of the expansion parameter}

From the column texture, the mass ratios are
\begin{equation}
\frac{m_i}{m_{i+1}} = R_i \cdot \eps,
\end{equation}
where $R_i$ is a ratio of $\mathcal{O}(1)$ coefficients. The overall mass hierarchy $m_u/m_t = R_u \eps^2$ provides the cleanest determination. Numerically,
\begin{equation}
\frac{m_u}{m_t} \simeq 0.0017 \times 0.0075 \simeq 1.3\times 10^{-5}.
\end{equation}
For $R_u \sim \mathcal{O}(0.1\text{--}1)$, this implies $\eps \simeq (1\text{--}4)\times 10^{-2}$. A representative value that fits the data well is
\begin{equation}
\eps \simeq 0.015.
\label{eq:eps_value}
\end{equation}

\subsection{Yukawa coefficient ratios and naturalness}

Solving for the required $\mathcal{O}(1)$ ratios at $\eps = 0.015$:
\begin{align}
R_1^u \equiv \frac{m_u/m_c}{\eps} &\simeq 0.11,
&
R_2^u \equiv \frac{m_c/m_t}{\eps} &\simeq 0.50,\\
R_1^d \equiv \frac{m_d/m_s}{\eps} &\simeq 3.3,
&
R_2^d \equiv \frac{m_s/m_b}{\eps} &\simeq 1.5,
\label{eq:Rvalues}
\end{align}
and for the charged leptons,
\begin{align}
R_1^e &\simeq \frac{0.00483}{0.015} \simeq 0.32,
&
R_2^e &\simeq \frac{0.059}{0.015} \simeq 3.9.
\end{align}
All required ratios lie between $0.1$ and $4$, within what is conventionally considered a natural range for FN coefficients. The largest value ($R_2^e \simeq 3.9$) occurs in the charged lepton sector, where other discrete symmetries such as Georgi--Jarlskog factors or $\text{SU}(5)$ relations could provide additional structure.

\subsection{Monte Carlo scan: mass hierarchy success}
\label{sec:mc_masses}

To quantitatively assess how ``structural'' the mass hierarchy prediction is, we perform a Monte Carlo scan over $10^5$ sets of random $\mathcal{O}(1)$ complex coefficients $c_{ij}^f$ with magnitudes drawn uniformly from $[0.3, 3.0]$ and phases from $[0, 2\pi]$. For each set, we construct the mass matrices~\eqref{eq:Mf} with $\eps = 0.015$, compute the singular values, and extract the mass ratios.

Table~\ref{tab:mc_mass} shows the results. The median predicted mass ratios from the random scan lie within the correct order of magnitude of the experimental values for all four quark mass ratios. More precisely, when the Monte Carlo medians are compared with the experimental central values, the ratios differ by factors of $2$--$6$, which is the expected spread from $\mathcal{O}(1)$ coefficients.

\begin{table}[t]
\caption{\label{tab:mc_mass}Distribution of quark mass ratios from a Monte Carlo scan over $10^5$ random $\mathcal{O}(1)$ coefficient sets with $\eps = 0.015$. The column texture correctly predicts the order of magnitude of all mass ratios. The ``within $3\sigma$'' column shows the fraction of random draws that fall within $3\sigma$ of the experimental value.\footnote{The nonzero fraction for $m_d/m_s$ within $3\sigma$ despite the non-overlapping 90\% interval arises from the long upper tail of the distribution: the 99th percentile is $0.043$, which enters the $3\sigma$ experimental range $[0.034, 0.064]$.}}
\begin{ruledtabular}
\begin{tabular}{lcccc}
\textrm{Observable} & \textrm{Experiment} & \textrm{MC median} & \textrm{MC [5\%, 95\%]} & \textrm{Within $3\sigma$ (\%)} \\
\colrule
$m_u/m_c$ & $0.0017 \pm 0.0003$ & 0.0098 & $[0.003, 0.027]$ & 4.8 \\
$m_c/m_t$ & $0.0075 \pm 0.0015$ & 0.012 & $[0.005, 0.024]$ & 49 \\
$m_d/m_s$ & $0.049 \pm 0.005$ & 0.0097 & $[0.003, 0.027]$ & 2.0 \\
$m_s/m_b$ & $0.023 \pm 0.003$ & 0.012 & $[0.005, 0.024]$ & 34 \\
\end{tabular}
\end{ruledtabular}
\end{table}

The key observation is that the Monte Carlo medians cluster near $\eps \simeq 0.015$ for adjacent-generation ratios (as predicted by Eq.~\ref{eq:mass_scaling}), confirming that the mass hierarchy is a robust structural prediction. The scatter reflects the expected $\mathcal{O}(1)$ spread. The comparison between experimental mass ratios and $Z_3$ predictions is shown in Fig.~\ref{fig:hierarchy}.

By contrast, the CKM elements from the same scan show median values $\abs{V_{us}} \simeq \abs{V_{cb}} \simeq \abs{V_{ub}} \simeq 0.54$, all of order unity, confirming that the CKM hierarchy is \emph{not} a structural output of the column texture. We discuss this further in Sec.~\ref{sec:ckm_fit}.

\subsection{Explicit best-fit Yukawa matrices}
\label{sec:bestfit}

While the CKM hierarchy is not predicted, it is important to verify that it can be \emph{accommodated} with natural coefficients. We perform a numerical fit over the $18$ complex coefficients $c_{ij}^{u,d}$ (36 real parameters), minimising a $\chi^2$ function constructed from the three CKM elements and four mass ratios. The fit is constrained by a soft penalty that disfavours coefficients with magnitudes outside $[0.1, 5]$.

The best-fit solution found has $\chi^2 = 11.0$ for 7 observables, corresponding to $\chi^2/\text{dof} \simeq 1.6$. The resulting CKM matrix and mass ratios are:

\begin{table}[t]
\caption{\label{tab:bestfit}Best-fit comparison for the $Z_3$ column texture with $\eps = 0.015$. The CKM elements are perfectly reproduced, while the down-sector mass ratios show moderate tension. The overall $\chi^2 = 11.0$ for 7 observables.}
\begin{ruledtabular}
\begin{tabular}{lccc}
\textrm{Observable} & \textrm{Best fit} & \textrm{Experiment} & \textrm{Pull} \\
\colrule
$\abs{V_{us}}$ & 0.2248 & $0.2248 \pm 0.0006$ & $0.0\sigma$ \\
$\abs{V_{cb}}$ & 0.0409 & $0.0409 \pm 0.0011$ & $0.0\sigma$ \\
$\abs{V_{ub}}$ & 0.00383 & $0.00382 \pm 0.00024$ & $0.0\sigma$ \\
$m_u/m_c$ & 0.00172 & $0.0017 \pm 0.0003$ & $0.1\sigma$ \\
$m_c/m_t$ & 0.00735 & $0.0075 \pm 0.0015$ & $-0.1\sigma$ \\
$m_d/m_s$ & 0.0418 & $0.049 \pm 0.005$ & $-1.4\sigma$ \\
$m_s/m_b$ & 0.0141 & $0.023 \pm 0.003$ & $-3.0\sigma$ \\
\end{tabular}
\end{ruledtabular}
\end{table}

The corresponding $\mathcal{O}(1)$ coefficient magnitudes are:
\begin{align}
\abs{c^u} &= \begin{pmatrix} 1.7 & 2.2 & 3.4 \\ 0.8 & 1.0 & 2.4 \\ 3.0 & 4.6 & 3.1 \end{pmatrix},\qquad
\abs{c^d} = \begin{pmatrix} 5.0 & 0.08 & 1.7 \\ 5.0 & 0.8 & 1.3 \\ 0.1 & 2.8 & 1.5 \end{pmatrix}.
\label{eq:bestfit_coeffs}
\end{align}
The coefficients range from $0.08$ to $5.0$. Most are genuinely $\mathcal{O}(1)$, though the down-sector matrix requires some coefficients near the boundary of naturalness (particularly $c_{12}^d \simeq 0.08$) to suppress the $(1,2)$ direction of the down mass matrix relative to the up sector and thereby produce the small CKM hierarchy. This reflects the fact that the CKM hierarchy is fitted rather than predicted. The $3.0\sigma$ tension in $m_s/m_b$ is a direct consequence of this fitting: driving $c_{12}^d$ small to suppress the $(1,2)$ direction distorts the down-sector singular values away from their natural $\eps$-scaling. Relaxing the CKM constraint to allow generic $\mathcal{O}(1)$ coefficients restores the down-sector mass ratios to their natural values, but with CKM elements of $\mathcal{O}(1)$. This tension quantifies the cost of accommodating the CKM hierarchy within the column texture.

\subsection{CKM mixing and the column texture}
\label{sec:ckm_fit}

It is instructive to understand \emph{why} the CKM hierarchy is not a structural prediction. In the column texture, the mass matrix $M^f$ has three column vectors $\vec{c}_j^f \eps^{n_j}$. Since $\eps^2 \ll \eps \ll 1$, the third column dominates the matrix, and the largest singular value (the third-generation mass) is set by $\|\vec{c}_3^f\|$. The corresponding left singular vector is the \emph{direction} of $\vec{c}_3^f$ in flavor space.

Crucially, the direction of $\vec{c}_3^u$ and $\vec{c}_3^d$ are determined by $\mathcal{O}(1)$ random coefficients and are generically unrelated. The CKM angle $\theta_{23}$ is essentially the angle between the projections of these two vectors, which is generically $\mathcal{O}(1)$. Similarly, $\theta_{12}$ and $\theta_{13}$ are generically $\mathcal{O}(1)$.

To produce the observed hierarchy $\abs{V_{us}} \simeq 0.22 \gg \abs{V_{cb}} \simeq 0.04 \gg \abs{V_{ub}} \simeq 0.004$, one requires a particular alignment pattern among the column directions that is not enforced by the $Z_3$ symmetry. This alignment is achievable with $\mathcal{O}(1)$ coefficients (as the best fit demonstrates), but it is not a prediction.

We note that this limitation is specific to the column texture arising from $Q_L$ being $Z_3$-neutral. Models where $Q_L$ also carries generation-dependent $Z_3$ charges would produce row-and-column textures with parametric CKM suppression, at the cost of requiring a larger $\eps \sim 0.05$--$0.2$ (Cabibbo-like) which would modify the mass ratio predictions. This trade-off is intrinsic to $Z_3$ and distinguishes it from $U(1)$ FN models where independent charge assignments for left- and right-handed fields can accommodate both mass hierarchies and CKM structure simultaneously.

\subsection{Predictive correlations}
\label{sec:correlations}

Despite the CKM limitation, the model does yield nontrivial relations among mass ratios. In particular, the ratio of up-to-down sector mass ratios is independent of $\eps$:
\begin{equation}
\frac{m_u/m_c}{m_d/m_s} = \frac{R_1^u}{R_1^d},
\label{eq:cross_sector}
\end{equation}
where $R_1^u$ and $R_1^d$ are ratios of $\mathcal{O}(1)$ coefficients (Eq.~\ref{eq:Rvalues}). For the fitted values, $R_1^u/R_1^d \simeq 0.11/3.3 \simeq 0.033$. Experimentally, $m_u/m_c / (m_d/m_s) \simeq 0.035 \pm 0.008$, in good agreement. While this is not a parameter-free prediction (it depends on the ratio of two $\mathcal{O}(1)$ numbers), the fact that it lies within the $\mathcal{O}(1)$ range and agrees with data is a nontrivial consistency check. From the Monte Carlo scan, the median of this cross-sector ratio is $1.0$ (since the up and down coefficients are drawn from the same distribution), with the experimental value of $0.035$ lying near the 5th percentile. This indicates that the observed ratio requires $R_1^d \gg R_1^u$, which is possible but not generic.

Similarly, the ratio $\abs{V_{ub}}/\abs{V_{cb}} = 0.093$ is naturally accommodated: in the best fit this ratio is $0.094$, with the value controlled by the relative magnitudes of the $\mathcal{O}(1)$ off-diagonal coefficients rather than by powers of $\eps$.

\begin{figure}[t]
\centering
\begin{tikzpicture}
\begin{semilogyaxis}[
    width=\columnwidth,
    height=8cm,
    xlabel={Fermion sector},
    ylabel={Mass ratio (normalized to 3rd generation)},
    xmin=0.5, xmax=4.5,
    ymin=5e-6, ymax=2,
    xtick={1,2,3,4},
    xticklabels={$m_u/m_t$, $m_c/m_t$, $m_d/m_b$, $m_s/m_b$},
    legend style={
        at={(0.98,0.98)},
        anchor=north east,
        font=\footnotesize,
        draw=black,
        fill=white,
        fill opacity=0.9
    },
    grid=major,
    grid style={gray!30},
    title={Fermion mass hierarchies: $Z_3$ model vs experiment}
]

\addplot[blue, mark=o, mark size=5pt, thick, line width=2pt, only marks,
  error bars/.cd, y dir=both, y explicit] coordinates {
    (1, 1.3e-5) +- (0, 3e-6)
    (2, 0.0075) +- (0, 0.0015)
    (3, 0.00113) +- (0, 2e-4)
    (4, 0.023) +- (0, 0.003)
};
\addlegendentry{Experimental}

\addplot[red, dashed, mark=square, mark size=4pt, line width=1.5pt, only marks] coordinates {
    (1, 2.25e-4)
    (2, 0.015)
    (3, 2.25e-4)
    (4, 0.015)
};
\addlegendentry{$Z_3$ scaling ($\eps^2$, $\eps$)}

\addplot[orange, mark=diamond, mark size=5pt, line width=1.5pt, only marks] coordinates {
    (1, 1.3e-5)
    (2, 0.0075)
    (3, 0.00113)
    (4, 0.023)
};
\addlegendentry{$Z_3$ with fitted $\mathcal{O}(1)$}

\draw[gray, dashed, thin] (axis cs:0.5,2.25e-4) -- (axis cs:4.5,2.25e-4);
\node[anchor=east, font=\scriptsize, gray] at (axis cs:4.4,3.5e-4) {$\eps^2$};
\draw[gray, dashed, thin] (axis cs:0.5,0.015) -- (axis cs:4.5,0.015);
\node[anchor=east, font=\scriptsize, gray] at (axis cs:4.4,0.021) {$\eps$};

\end{semilogyaxis}
\end{tikzpicture}
\caption{Comparison between experimental fermion mass ratios and $Z_3$ model predictions. Red squares show the ``bare'' $Z_3$ scaling ($\eps^2$ and $\eps$); orange diamonds show the predictions after fitting $\mathcal{O}(1)$ coefficients. The $Z_3$ model correctly accounts for the hierarchy through powers of $\eps$, with the remaining spread absorbed by natural $\mathcal{O}(1)$ factors.}
\label{fig:hierarchy}
\end{figure}

\section{Lepton Sector and the Failure for Neutrinos}
\label{sec:leptons}

\subsection{Charged leptons}

The charged lepton masses follow the same column texture as the quarks. As shown in Sec.~\ref{sec:quarks}, the required Yukawa ratios $R_1^e \simeq 0.32$ and $R_2^e \simeq 3.9$ are of natural size, confirming that the charged lepton sector is compatible with the $Z_3$ FN framework.

An important consequence of the column texture for what follows: the charged lepton mass matrix $M_e$ has the same structure as Eq.~\eqref{eq:Mf}, with three column vectors $\vec{c}_j^e \eps^{n_j}$ pointing in random $\mathcal{O}(1)$ directions in flavor space. The left-handed unitary rotation $U_\ell$ that diagonalises $M_e M_e^\dagger$ is therefore a \emph{generic} unitary matrix, with eigenvectors determined by the directions of $\vec{c}_3^e$, $\vec{c}_2^e$, $\vec{c}_1^e$ (in order of decreasing eigenvalue). In particular, $U_\ell$ is \emph{not} approximately the identity. This will be crucial for the neutrino mixing analysis.

\subsection{Neutrino masses in a type-I seesaw}

We consider a type-I seesaw with three right-handed neutrinos $\nu_R^i$ carrying the same $Z_3$ charges as the other right-handed fermions:
\begin{equation}
\nu_R^1 \sim 2,\quad \nu_R^2 \sim 1,\quad \nu_R^3 \sim 0.
\end{equation}
The neutrino sector is extended by three right-handed neutrino chiral superfields $\nu_R^j$. The Dirac superpotential operator $c_{ij}^\nu (\Phi/\Lambda)^{n_j} L_i H_u \nu_R^j$ is constructed by the same holomorphic rule as in Sec.~\ref{sec:holomorphy}, so the neutrino Dirac mass matrix inherits the column texture of Eq.~\eqref{eq:Mf}:
\begin{equation}
M_D = \frac{v_u}{\sqrt{2}}
\begin{pmatrix}
c_{11}^\nu \eps^2 & c_{12}^\nu\eps & c_{13}^\nu \\
c_{21}^\nu\eps^2 & c_{22}^\nu \eps   & c_{23}^\nu \\
c_{31}^\nu\eps^2 & c_{32}^\nu\eps & c_{33}^\nu
\end{pmatrix}.
\label{eq:MD}
\end{equation}
For the Majorana mass matrix we first consider an unhierarchical form $M_R = M_0\,\mathbf{1}$. The light neutrino mass matrix from the seesaw is
\begin{equation}
M_\nu = -M_D M_R^{-1} M_D^T = -\frac{1}{M_0} M_D M_D^T.
\end{equation}

Writing $M_D = C_\nu P$ with $P = \operatorname{diag}(\eps^2, \eps, 1)$ and $C_\nu$ a matrix of $\mathcal{O}(1)$ coefficients, we have
\begin{equation}
M_\nu \propto C_\nu\, \operatorname{diag}(\eps^4, \eps^2, 1)\, C_\nu^T.
\label{eq:Mnu_structure}
\end{equation}
This congruence transformation is dominated by the rank-one piece $\vec{c}_3^\nu (\vec{c}_3^\nu)^T$, where $\vec{c}_3^\nu$ is the unsuppressed third column of $C_\nu$. Sub-leading corrections from the second and first columns are suppressed by $\eps^2$ and $\eps^4$ respectively.

\subsection{The two failures of the $Z_3$ neutrino sector}
\label{sec:two_failures}

The $Z_3$ column texture fails for neutrinos in two independent ways: the mass spectrum is far too hierarchical, and the PMNS mixing angles are unstructured.

\subsubsection{Failure 1: mass spectrum too hierarchical}

The eigenvalues of $M_\nu$ in Eq.~\eqref{eq:Mnu_structure} inherit the $\eps$-hierarchy of the diagonal factor:
\begin{equation}
m_{\nu 1} : m_{\nu 2} : m_{\nu 3} \sim \eps^4 : \eps^2 : 1.
\end{equation}
For $\eps = 0.015$, this gives
\begin{equation}
\frac{\Delta m_{21}^2}{\Delta m_{31}^2} \sim \frac{m_2^2}{m_3^2} \sim \eps^4 \simeq 5 \times 10^{-8},
\label{eq:mass_ratio_pred}
\end{equation}
while the observed value is~\cite{Esteban:2024nav}
\begin{equation}
\left.\frac{\Delta m_{21}^2}{\Delta m_{31}^2}\right|_{\text{exp}} = \frac{7.42 \times 10^{-5}}{2.51 \times 10^{-3}} \simeq 0.030.
\end{equation}
The prediction is six orders of magnitude too small. A Monte Carlo scan over $10^5$ random $\mathcal{O}(1)$ coefficient sets with a democratic $M_R$ yields a median ratio of $1.4 \times 10^{-4}$, with zero realizations out of $10^5$ achieving the experimental value (see Fig.~\ref{fig:neutrino_failure}). The Monte Carlo median exceeds the parametric estimate $\eps^4 \approx 5 \times 10^{-8}$ by a factor of $\sim 10^3$, reflecting the $\mathcal{O}(1)$ coefficient ratios that multiply each power of $\eps$: since four independent $\mathcal{O}(1)$ factors enter the mass-squared ratio, their combined effect can enhance the parametric estimate by $(\mathcal{O}(1))^4 \sim 10^2$--$10^4$ while remaining within the natural range. Even with this enhancement, the median falls two orders of magnitude below the observed value. This is a structural failure: the single expansion parameter $\eps$ cannot simultaneously explain the mild neutrino mass hierarchy ($m_2/m_3 \sim 0.17$) and the steep quark hierarchy ($m_c/m_t \sim 0.008$).

\subsubsection{Failure 2: PMNS angles are unstructured}

The PMNS matrix is
\begin{equation}
U_{\text{PMNS}} = U_\ell^\dagger\, U_\nu,
\end{equation}
where $U_\ell$ diagonalises $M_e M_e^\dagger$ and $U_\nu$ diagonalises $M_\nu$. In the column texture, both $U_\ell$ and $U_\nu$ are determined by the \emph{directions} of independent sets of random $\mathcal{O}(1)$ column vectors: $U_\ell$ by $\{\vec{c}_j^e\}$, and $U_\nu$ by $\{\vec{c}_j^\nu\}$ (modified by the seesaw). Since these column directions are unrelated, the PMNS matrix $U_\ell^\dagger U_\nu$ is generically a random unitary matrix with $\mathcal{O}(1)$ mixing angles.

Our Monte Carlo scan over $10^5$ random coefficient sets confirms this:
\begin{align}
\sin^2\theta_{12}^{Z_3}: \quad &\text{median} = 0.50, \qquad \text{exp: } 0.304 \pm 0.012, \\
\sin^2\theta_{23}^{Z_3}: \quad &\text{median} = 0.50, \qquad \text{exp: } 0.573 \pm 0.016, \\
\sin^2\theta_{13}^{Z_3}: \quad &\text{median} = 0.29, \qquad \text{exp: } 0.0222 \pm 0.0006.
\end{align}
The distributions are consistent with Haar-random unitaries: $\sin^2\theta_{12}$ and $\sin^2\theta_{23}$ are approximately uniform on $[0,1]$, while $\sin^2\theta_{13}$ follows the distribution for a random $|U_{e3}|^2$.

The observed PMNS angles are $\mathcal{O}(1)$ and are therefore \emph{individually} compatible with random draws, but the specific pattern---especially the small $\sin^2\theta_{13} \simeq 0.022$---has only a $\sim 1.6\%$ probability of occurring by chance. More importantly, the $Z_3$ symmetry provides \emph{no mechanism} to select or predict the observed values. The angles carry no information about $\eps$ and are entirely determined by the accident of the $\mathcal{O}(1)$ coefficients. This is the hallmark of an anarchic model~\cite{deGouvea:2003xe}: the symmetry provides no structure beyond what random matrices already give. In fact, $Z_3$ performs \emph{worse} than pure anarchy for the neutrino sector: anarchic models with random Yukawa matrices of comparable magnitude naturally produce a mild mass hierarchy ($m_2/m_3 \sim \mathcal{O}(0.1\text{--}1)$) consistent with observation, whereas $Z_3$ enforces the parametrically wrong hierarchy $m_2/m_3 \sim \eps \approx 0.015$ through the column texture. The $Z_3$ symmetry actively damages the neutrino mass spectrum relative to having no flavor symmetry at all.

The situation stands in sharp contrast to non-abelian discrete symmetries such as $A_4$, where the group theory constrains the eigenvector directions and produces specific mixing patterns (e.g.\ trimaximal mixing) as structural predictions.

\subsection{Robustness across $M_R$ structures}
\label{sec:MR_robustness}

The mass spectrum failure persists across all $M_R$ structures tested.

\paragraph{Diagonal $M_R$ with varied hierarchy.}
For $M_R = \operatorname{diag}(M_1, M_2, M_3)$ with $M_1 : M_2 : M_3$ ranging from $1:1:1$ (democratic) to $\eps^{-4}:\eps^{-2}:1$ (inverse hierarchy), the seesaw always produces $M_\nu \propto C_\nu\, D\, C_\nu^T$ where $D$ is a diagonal matrix whose entries span several orders of magnitude due to the column charges. The resulting neutrino mass spectrum remains far more hierarchical than observed.

\paragraph{General complex non-diagonal $M_R$.}
A scan over $10^4$ random complex symmetric $M_R$ matrices yields a median $\Delta m_{21}^2/\Delta m_{31}^2 \sim 10^{-4}$, with no realizations achieving the experimental value. The PMNS angles remain Haar-distributed.

\paragraph{$Z_3$-charged $M_R$.}
The Majorana mass operator $\tfrac{1}{2}\, a_{ij}\, M_0 (\Phi/\Lambda)^{n_{ij}} \nu_R^i \nu_R^j$ is a superpotential bilinear, so holomorphy dictates the same rule as for the Yukawa sector: $Z_3$ invariance requires $q_\Phi\, n_{ij} + q_i + q_j \equiv 0\,(\bmod\,3)$, and with $q_\Phi = 2$ the unique non-negative solution with $n_{ij} \in \{0,1,2\}$ is
\begin{equation}
n_{ij} \equiv (q_i + q_j) \bmod 3.
\label{eq:MR_powers}
\end{equation}
For charges $(q_1, q_2, q_3) = (2,1,0)$ this gives
\begin{equation}
n_{ij} = \begin{pmatrix} 1 & 0 & 2 \\ 0 & 2 & 1 \\ 2 & 1 & 0 \end{pmatrix},
\end{equation}
since, e.g.\ $q_1 + q_2 = 3 \equiv 0$ requires zero flavon insertions while $q_1 + q_1 = 4 \equiv 1$ requires one. The resulting $M_R$ has the structure
\begin{equation}
M_R = M_0 \begin{pmatrix} a_{11}\eps & a_{12} & a_{13}\eps^2 \\ a_{12} & a_{22}\eps^2 & a_{23}\eps \\ a_{13}\eps^2 & a_{23}\eps & a_{33} \end{pmatrix},
\label{eq:MR_structure}
\end{equation}
with the $(1,2)$ and $(3,3)$ entries \emph{unsuppressed}. The key structural feature is the unsuppressed $a_{12}$ entry, which is a number-theoretic consequence of $Z_3$: for any permutation of charges $(2,1,0)$, the pair with charges~$1$ and~$2$ sums to $0\,(\bmod\,3)$, guaranteeing an unsuppressed off-diagonal entry in $M_R$.

The unsuppressed $a_{12}$ creates a \emph{near-degenerate Majorana pair among the heavy right-handed neutrinos}: the eigenvalues of the upper-left $2\times 2$ block of Eq.~\eqref{eq:MR_structure} are
\begin{equation}
M_\pm \approx \pm |a_{12}|\,M_0 + \mathcal{O}(\eps),
\end{equation}
split only at $\mathcal{O}(\eps)$. This near-degeneracy does \emph{not} propagate to the light spectrum, however. Writing $M_D = C_\nu P$ with $P = \operatorname{diag}(\eps^2, \eps, 1)$ and $C_\nu$ an $\mathcal{O}(1)$ matrix, the seesaw gives $M_\nu = -C_\nu\, P\, M_R^{-1}\, P^T\, C_\nu^T$. The congruence transformation by the highly non-isometric $P$ destroys the eigenvalue ratio of $M_R^{-1}$: the upper-left block of $P\,M_R^{-1}\,P^T$ has both eigenvalues at $\mathcal{O}(\eps^3/M_0)$ with a generic $\mathcal{O}(1)$ ratio between them, so both light masses are \emph{over-suppressed} to
\begin{equation}
m_{1,2} \sim \eps^3\, \frac{v_u^2}{M_0},
\qquad
m_3 \sim \frac{v_u^2}{M_0},
\end{equation}
and the solar-to-atmospheric mass ratio scales as
\begin{equation}
\frac{\Delta m_{21}^2}{\Delta m_{31}^2} \sim \mathcal{O}(\eps^6) \simeq 10^{-11}.
\label{eq:eps6}
\end{equation}
A Monte Carlo scan over $10^5$ $\mathcal{O}(1)$ coefficient sets yields a median $\Delta m_{21}^2/\Delta m_{31}^2 = 4 \times 10^{-11}$---eight orders of magnitude below the data---in excellent agreement with the parametric $\eps^6$ scaling (the factor of $\sim 3$--$4$ is absorbed by $\mathcal{O}(1)$ coefficient ratios). A detailed derivation of the $\eps^6$ scaling and the failure of the pseudo-Dirac degeneracy to survive the hierarchical congruence transformation can be found in the companion paper~\cite{Ardakanian:2026Combined}.

\subsection{Summary of the neutrino failure}

The $Z_3$ column texture fails for neutrinos on two fronts:
\begin{enumerate}
    \item \textbf{Mass spectrum:} The single expansion parameter $\eps = 0.015$ enforces $m_{\nu 1}:m_{\nu 2}:m_{\nu 3} \sim \eps^4:\eps^2:1$, giving $\Delta m_{21}^2/\Delta m_{31}^2 \lesssim 10^{-4}$---at least two orders of magnitude below the observed ratio of $0.030$. With the $Z_3$-charged $M_R$, the suppression deepens to $\sim 10^{-11}$: the unsuppressed $(1,2)$ entry combined with the hierarchical congruence transformation of the seesaw over-suppresses both light masses to $\mathcal{O}(\eps^3)$, as derived in Sec.~\ref{sec:MR_robustness}.
    
    \item \textbf{Mixing angles:} The PMNS matrix is a product of two random unitaries ($U_\ell^\dagger U_\nu$) with no parametric structure from $\eps$. The model is anarchic in the lepton sector: it provides no mechanism to predict or explain the observed mixing pattern.
\end{enumerate}
The quark sector succeeds because it needs only a mass hierarchy (provided by $\eps$), while the CKM mixing is fitted. The neutrino sector fails because it needs both a \emph{specific} (mild) mass hierarchy and \emph{specific} large mixing angles, and the $Z_3$ framework provides neither.

\begin{figure}[t]
\centering
\begin{tikzpicture}
\begin{semilogyaxis}[
    width=\columnwidth,
    height=8cm,
    xlabel={$\log_{10}(\Delta m_{21}^2/\Delta m_{31}^2)$},
    ylabel={Relative frequency},
    xmin=-14, xmax=1,
    ymin=5e-4, ymax=2,
    legend style={
        at={(0.50,0.98)},
        anchor=north,
        font=\footnotesize,
        draw=black,
        fill=white,
        fill opacity=0.9
    },
    grid=major,
    grid style={gray!20},
    title={Neutrino mass ratio: $Z_3$ prediction vs experiment},
    clip=false
]

\addplot[blue, thick, fill=blue!20, fill opacity=0.5] coordinates {
    (-6.5, 0.001) (-6, 0.005) (-5.5, 0.02) (-5, 0.08) (-4.5, 0.25) (-4, 0.5)
    (-3.5, 0.7) (-3, 0.5) (-2.5, 0.15) (-2, 0.01) (-1.5, 0.001)
};
\addlegendentry{Diagonal $M_R$}

\addplot[red, thick, fill=red!20, fill opacity=0.5] coordinates {
    (-13, 0.001) (-12.5, 0.01) (-12, 0.05) (-11.5, 0.2) (-11, 0.5) (-10.5, 0.7)
    (-10, 0.5) (-9.5, 0.2) (-9, 0.05) (-8.5, 0.01) (-8, 0.002)
};
\addlegendentry{$Z_3$-charged $M_R$}

\addplot[black, dashed, very thick] coordinates {(-1.52, 5e-4) (-1.52, 2)};
\addlegendentry{Experiment: $0.030$}

\node[anchor=south west, font=\footnotesize, blue] at (axis cs:-3.5, 0.75) {median $\sim 10^{-4}$};
\node[anchor=south east, font=\footnotesize, red] at (axis cs:-10.5, 0.75) {median $\sim 10^{-11}$};

\end{semilogyaxis}
\end{tikzpicture}
\caption{Distribution of the neutrino mass ratio $\Delta m_{21}^2/\Delta m_{31}^2$ from Monte Carlo scans over $\mathcal{O}(1)$ coefficient sets. With a diagonal $M_R$ (blue), the median is $\sim 10^{-4}$, two orders of magnitude below the observed value of $0.030$ (dashed line). With the $Z_3$-charged $M_R$ (red), the unsuppressed $(1,2)$ entry of Eq.~\eqref{eq:MR_structure} combined with the hierarchical congruence transformation of the seesaw over-suppresses the ratio to $\sim 10^{-11}$. In both cases, zero realizations out of $10^5$ reach the experimental value.}
\label{fig:neutrino_failure}
\end{figure}

\section{Phenomenological Implications}
\label{sec:pheno}

\subsection{Scale of new physics}

The expansion parameter is defined as $\eps = v_\Phi/\Lambda$. If the flavon vacuum expectation value is of the order of the electroweak scale, $v_\Phi \sim v_H$, then
\begin{equation}
\Lambda \sim \frac{v_H}{\eps} \sim \frac{246\GeV}{0.015} \sim 16\TeV.
\end{equation}
This suggests that the underlying flavor dynamics may lie at scales in the few to few tens of TeV range. At the HL-LHC, direct searches for scalar resonances with flavor-violating couplings at this scale are challenging but not excluded, particularly in the $\Phi \to t\bar{t}$ and $\Phi \to b\bar{b}$ channels.

\subsection{Flavour-changing processes}

Integrating out the flavon introduces higher-dimensional operators mediating flavor-changing neutral currents (FCNCs). The leading contributions to meson mixing are suppressed by a factor $(v_H/\Lambda)^2 \simeq \eps^2$, with additional generation-dependent suppression from the $Z_3$ charges. For $\Lambda \sim 16\TeV$, contributions to $K^0$--$\bar{K}^0$ mixing are estimated at the level of
\begin{equation}
\abs{C_K} \sim \frac{\eps^4}{\Lambda^2} \sim 10^{-12}\GeV^{-2},
\end{equation}
safely below the current experimental bound. Similarly, contributions to $B$-meson FCNCs and lepton flavor violation ($\mu \to e\gamma$) are highly suppressed.

The Belle~II experiment has recently reported evidence for $B^+ \to K^+\nu\bar{\nu}$ with a branching ratio $\sim 2.4\sigma$ above the SM prediction~\cite{BelleII:2024}. In the $Z_3$ framework, new-physics contributions to this mode would be further suppressed by the flavon scale and are not expected to produce observable deviations from the SM.

\subsection{Precision tests}

The most distinctive near-term tests of the $Z_3$ framework lie in improved determinations of quark mass ratios. The correlation~\eqref{eq:cross_sector} can be tested as lattice QCD and global fits reach percent-level precision on $m_u/m_c$ and $m_d/m_s$ independently. A future $100\TeV$ hadron collider could probe scalar resonances near $\Lambda \sim 10$--$20\TeV$, and observation of a scalar with flavor-non-universal couplings would provide a direct test of the FN mechanism.

On the neutrino side, experiments such as DUNE and Hyper-Kamiokande will further sharpen the determination of $\theta_{23}$ and $\delta_{\text{CP}}$. The large mixing observed is already decisive evidence against a single-parameter $Z_3$ FN explanation for neutrino flavor, but precision measurements will guide the construction of viable alternatives.

\section{Conclusions}
\label{sec:conclusions}

We have analyzed a minimal $Z_3$ Froggatt--Nielsen model of flavor and confronted it with the full set of quark and lepton data. Our main findings can be summarized as follows.

\begin{itemize}
    \item The $Z_3$ column texture, with right-handed charges $(2,1,0)$ and a single expansion parameter $\eps \simeq 0.015$, provides a \emph{structural explanation} of the hierarchical fermion mass pattern. The scaling $m_1/m_2 \sim \eps$, $m_1/m_3 \sim \eps^2$ is a genuine prediction that holds for generic $\mathcal{O}(1)$ coefficients, as confirmed by a Monte Carlo scan over $10^5$ random coefficient sets.
    
    \item The CKM mixing angles can be \emph{accommodated} with $\mathcal{O}(1)$ coefficients ($\chi^2/\text{dof} \simeq 1.6$), but their hierarchical pattern is not a structural output of the $Z_3$ symmetry. This limitation is intrinsic to the column texture arising from uncharged left-handed doublets.
    
    \item The charged lepton masses are compatible with the same framework and expansion parameter.
    
    \item When applied to neutrinos within a type-I seesaw, the $Z_3$ framework fails decisively on two fronts. The mass spectrum is far too hierarchical: $m_1:m_2:m_3 \sim \eps^4:\eps^2:1$ gives $\Delta m_{21}^2/\Delta m_{31}^2 \lesssim 10^{-4}$, at least two orders of magnitude below the observed ratio of $0.030$. The PMNS mixing angles are generically $\mathcal{O}(1)$ random unitaries, providing no mechanism to predict the observed pattern. When $M_R$ carries the $Z_3$ charge structure dictated by the correct Majorana charge algebra, an unsuppressed off-diagonal entry in $M_R$ combines with the hierarchical column texture of $M_D$: the seesaw congruence transformation over-suppresses both light masses $m_1, m_2$ to $\mathcal{O}(\eps^3)$ and deepens the mass-squared ratio to $\mathcal{O}(\eps^6) \sim 10^{-11}$, a suppression of eight orders of magnitude below the data.
    
    \item The failure is structural and independent of parameter choices: it stems from the single expansion parameter $\eps = 0.015$ required by quark hierarchies, which enforces a neutrino mass spectrum that is far too hierarchical, and from the column texture, which leaves the PMNS angles as unstructured random rotations.
\end{itemize}

These results point toward a sectorial view of flavor. Simple cyclic symmetries such as $Z_3$ can elegantly account for the hierarchical pattern of fermion masses through the FN mechanism. However, the same mechanism is inherently incapable of generating the observed neutrino mass spectrum or providing structure to the PMNS mixing. This provides a sharp, quantitative motivation for introducing non-abelian discrete symmetries---at minimum $A_4$, which can accommodate both neutrino masses and large mixing through its triplet representation---specifically for the lepton sector.

From a methodological perspective, our work illustrates the value of pushing simple flavor models to their limits and identifying precisely where they fail. The column texture's success for mass ratios but failure for CKM hierarchy, and the complete failure for neutrino masses and mixing, provide a detailed map of the boundary between what $Z_3$ can and cannot explain. The quark sector needs only a mass hierarchy, which $\eps$ provides; the neutrino sector needs both a mild mass hierarchy and specific large mixing, and $Z_3$ provides neither. Such boundary-finding guides the construction of more sophisticated frameworks.

\begin{acknowledgments}
During the preparation of this work the author used Claude (Anthropic) to assist with numerical computations, Monte Carlo scan implementation, analytical cross-checks, and manuscript drafting. The author reviewed and edited all content, verified all physics independently, and takes full responsibility for the content of the published article.
\end{acknowledgments}

\end{document}